\documentstyle[aps,12pt]{revtex}
\begin{document}
\tightenlines
\title{\Large Classical  solutions from quantum regime for barotropic FRW 
model}
\author{J. Socorro} 
\address{Instituto de F\'{\i}sica de la Universidad de Guanajuato,\\
Apartado Postal E-143, C.P. 37150, Le\'on, Guanajuato, M\'exico\\
}

\maketitle
\widetext

\begin{abstract}
The quantization of gravity coupled to barotropic perfect fluid as ma\-tter 
field and cosmological constant is made and the wave
function can be determined for any $\kappa$ in the FRW minisuperspace model.
The meaning of the existence of the classical solution is discussed in the
WKB semiclassical approximation.\\
\\
{\it Keywords}: Classical and quantum  exact solutions.\\
\end{abstract}

\vspace{0.3cm}
\noindent {PACS numbers: 04.20.Jb; 04.40.Nr;  98.80.Hw}

\narrowtext

\section{Introduction}

It is a general belief that the existence of an initial singularity can be
remove through the employment of a quantum theory of gravity. However, there
is no consistent theory of gravity until now, and in this sense the problem
of the initial singularity remains of actuality.

It is well known that is possible to construct a quantum model for the 
universe as a whole, through the Wheeler-DeWitt (WDW) equation, based
in the ADM decomposition of the gravity sector, which leads to a hamiltonian
approach of  general relativity, from which a canonical quantization procedure
can be applied (Halliwel, 1991). Moreover, in the hamiltonian formalism, the
notion of time is lost (Isham, 1992), but there are some proposals by which
this notion of time can be recovered (Schutz, 1970, 1971) coupling 
of the
gravity sector to a perfect fluid. In this scheme, called Schutz's formalism,
a quantization procedure is possible, and the canonical momentum associated
with the perfect fluid appear linearly in the WDW equation, permiting to
rewrite this equation in the form a Schr\"odinger equation with a time
coordinate  associated with the matter field.

Recently, a convincing quantum gravity origin for inflation should 
however be proved.
According to Bojowald (Bojowald, 2002) a non-perturbative approach would 
produce the most reliable
answer as to whether or not inflation can be derived from quantum gravity. 
At the less ambitious level of minisuperspace quantum cosmology the
generality of inflation as well as the quantum creation of the universe
should be treated by analyzing the Wheeler-DeWitt equation and its 
physical solutions,
the so-called wave functions of the universe (Gibbons and Grishchuk, 1989).

In this way, the behaviour of the scale factor may be determined in 
two different forms from the quantum regime: when the wave function is time 
dependent, we can
calculate the expectation value of the scale factor, in the spirit
of the many worlds interpretation of quantum mechanics ( Tipler, 1986)
\begin{equation}\rm
<A>_t= \frac{\int_0^\infty {\cal W}(A) \, \Psi^*(A,t) \,A(t)\, \Psi(A,t)\, dA}
{\int_0^\infty {\cal W}(A) \, \Psi^*(A,t) \, \Psi(A,t) \, dA},
\end{equation}
where $\rm {\cal W}(A)$ is a weight function that normalizes the 
expectation value. The other way is to apply  the  WKB semiclassical
approximation, in which case the system follows a real trayectory given by the
equation
\begin{equation}
\rm \Pi_q= \frac{\partial \Phi}{\partial q},
\label{wkb}
\end{equation}
where the index q designates one of the degree of freedom of the
system, and $\Phi$ is the phase of the wave function when is written as
\begin{equation}
\rm \Psi = W \, e^{i\Phi},
\end{equation}
the functions W and $\Phi$ are real functions. 
We follow this last procedure to obtain the classical behaviour for
the scale factor, using the FRW model with a barotropic 
perfect fluid and cosmological constant.

The work is organized as follow. In next section we describe the quantum
model with the solution for any $\rm \kappa$ case in the minisuperspace, 
considering
a barotropic perfect fluid as matter field including the cosmological
constant. In section III, we present
the classical evolution for the scale factor derived following
the semiclassical WKB procedure. Section IV is devoted to conclusions.

\section{Quantum model }
The total lagrangian  for a barotropic perfect fluid coupled to gravity using  
the FRW geometry with the classical cosmological constant term  
is given by
\begin{equation}
{\cal L}_{total}= {\cal L}_{geom} + {\cal L}_{matter}.
\label{lagrangian}
\end{equation}
namely
\begin{eqnarray}
 {\cal L}_{geom} &=& {\rm \sqrt{-^{(4)}g} \, \left [ R - 2 \Lambda \right ]} = 
{\rm - \frac{6A^2}{N}\frac{d^2 A}{dt^2} 
 -\frac{6A}{N} \left(\frac{dA}{dt}\right)^2 }
+{\rm  \frac{6A^2}{N^2} \frac{dA}{dt} \frac{dN}{dt} - 6 \kappa N A
-2N \Lambda A^3 }\nonumber\\ 
&=&{\rm \frac{d}{dt} \left( \frac{-6A^2 \dot A}{N} \right) + \frac{6A}{N}
\left(\frac{dA}{dt}\right)^2 -  6 \kappa N A- 2N \Lambda A^3 },
\end{eqnarray}
and the lagrangian matter density (Ryan, 1972; Pazos, 2000)
\begin{eqnarray}
 {\cal L}_{matter}&=&{\rm \sqrt{-^{(4)}g} 
\left[ 16 \pi G N  \rho \left\{ (\gamma +1) 
\left(1+g^{km} \, U_k\, U_m\right)^{\frac{1}{2} } 
 - \gamma  \left(1+g^{km} \, U_k\, U_m\right)^{-\frac{1}{2} } 
\right\} \right.} \nonumber\\
&& \left.  - {\rm 16\pi G \rho (\gamma +1) U_m N^m } \right].
\end{eqnarray}
These results are obtained by employing the ADM form of the FRW metric
\begin{equation}
\rm ds^2 =
- N^2\,dt^2 +A^2 \left[\frac{dr^2}{1-\kappa r^2}+r^2\left(d\theta^2 +
sin^2\theta d\phi^2\right)\right],
\label{metrica}
\end{equation}
where $\rm N$ is the lapse function, $\rm A$ is the scale factor of the model,
and $\kappa$ is the curvature index of the universe
( $\rm \kappa=0,+1,-1$ plane, close and open, respectively).

We also make use of the usual perfect fluid energy-momentum tensor 
\begin{equation}
\rm T_{\mu\nu} = pg_{\mu\nu} +\left(p+\rho\right)U_\mu U_\nu~,  
\end{equation} 
where $\rm p$, $\rho$, $\rm U_\mu$ are the pressure, energy density and the
four-velocity of the cosmological fluid, respectively; 
and using the barotropic relationship $\rm  p =\gamma \rho$, 
$\gamma =$ constant, 
we have the solution for the energy density as a function of the scale 
factor of the FRW universe,
in the usual way
\begin{equation}
\rm \rho = \frac{M_\gamma}{A^{3 \left(\gamma+1\right)}},
\end{equation}
where $\rm M_\gamma$ is an integration constant.
In ${\cal L}_{matter}$ we also choose the comoving fluid  
(three-velocity $\rm U_k=0$), and the gauge $\rm N^k=0$, obtaining
\begin{equation}
{\cal L}_{matter} = \rm  16\pi G NM_\gamma A^{-3 \gamma}.
\end{equation}
Thus, the total lagrangian density has the following form
\begin{equation}
{\cal L}_{tot} = \rm \frac{d}{dt} \left( \frac{-6A^2 \dot A}{N} \right) +
 \frac{6A}{N}\left(\frac{dA}{dt}\right)^2
 -  6 \kappa N A -2N \Lambda A^3 + 16\pi G N M_\gamma A^{-3 \gamma}.
\end{equation}
Following the well-known procedure to get the canonical Hamiltonian 
function, we define the canonical momentum conjugate to the generalized 
coordinate A (scale factor) as 
\begin{equation}
\rm \Pi_A \equiv \frac{\partial L}{\partial \dot A},
\label{momento}
\end{equation}

\begin{equation}
\rm L =  \Pi_A\dot A 
- N {\cal H}=\Pi_A\dot A- N\left[\frac{\Pi ^2_A}{24A} +6\kappa A
+2 \Lambda A^3  -
16\pi G M_\gamma A^{-3\gamma}\right],
\label{lagrangiano}
\end{equation}
where
\begin{equation}
{\cal H} =\rm \frac{\Pi ^2_A}{24A} +6\kappa A +2 \Lambda A^3 
- 16 \pi G M_\gamma A^{-3\gamma}.
\label{hamiltoniana}
\end{equation}
Performing the variation of  (\ref{lagrangiano}) 
with respect to N,  $\rm \frac{\delta L}{\delta N}=0 $, implies the 
well-known result ${\cal H} =0$. Imposing the quantization condition and
applying this hamiltonian to the wave function $\Psi$, we
obtain the WDW equation in the minisuperspace 

\begin{equation}
\rm \hat {\cal H}  |\Psi \rangle = \frac{1}{24A}\left[-\frac{d^2}{dA^2} 
+144\kappa A^2 
+48 \Lambda A^4 -384\pi G  M_\gamma A^{-3\gamma +1}\right] |\Psi
\rangle=0.
\label{hamiltonian1}
\end{equation}
Notice that in principle the order ambiguity in equation 
(\ref{hamiltonian1}) should be taken into account. This is quite a 
difficult problem to be treated in all its generality, since the 
hamiltonian operator in (\ref{hamiltonian1}) must be written in 
a very general form in order to take into account all possible order,
but, at least in the minimal case in which (Hartle and Hawking, 1983) 
(there are other possibilities depending on 
different considerations on the operators, see the works of 
Christodoulakis and  Zanelli,  (Christodoulakis and Zanelli, 1984)
or Lidsey and Moniz (Lidsey and Moniz, 2000) 
\begin{equation}
\rm A^{-1} \frac{d^2 \Psi }{d A^2} \rightarrow A^{-1+p} \frac{d }{d A} A^{-p} 
\frac{d \Psi}{d A}= 
A^{-1}\left( \frac{d^2 \Psi}{d A^2} - p A^{-1} \frac{d \Psi}{dA} \right ),
\end{equation}
where the real parameter $\rm p$ measures the ambiguity in the factor ordering.
Therefore, the Wheeler-DeWitt equation can be written as follows
\begin{equation}
 - A\frac{d^2\Psi}{dA^2} + p \frac{d \Psi}{dA} - V(A)\Psi=0,
\label{WDW}
\end{equation}
where 
\begin{equation}
\rm V(A)=-48 \Lambda A^5 + 384 \pi G M_\gamma A^{-3\gamma + 2} 
- 144 \kappa A^3 .
\label{pote}
\end{equation}

In the following we discuss some of the quantum solutions of  
equation (\ref{WDW}) 
for particular values of the $\gamma$ parameter and the 
parameter p of factor ordering.
\subsection{Inflationary scenario}  
In the inflationary era, including the cosmological constant $\Lambda$, we 
choose  $\gamma=-1$. 
Notice that the WDW equation (\ref{WDW}) can be written as
\begin{equation}
 \rm A\frac{d^2\Psi}{dA^2} - p \frac{d \Psi}{dA} 
+ 144 A^3 \left( m^2 A^2 - \kappa
\right) \Psi=0, \label{21}
\end{equation}
where
\begin{equation}
\rm m^2= -\frac{\Lambda}{3} + \frac{8}{3}\pi G M_{-1} \, .
\label{hubble}
\end{equation} 
The latter relationship leads to three possible cases

\begin{enumerate}
\item{}  $\rm m^2 > 0$  \footnote{In general $\rm m^2$ is not a 
positive constant, see the corresponding definition (\ref{hubble}).}  

The differential equation for this subcase is

\begin{equation}
\rm A \Psi^{\prime \prime}_{-1} -Q \Psi^{\prime}_{-1} + 144\kappa A^3
\left( m^2 A^2 - \kappa  \right)  \Psi_{-1}=0
\end{equation}
which, after the substitutions $\rm v=m^2 A^2 - \kappa$, 
$\rm \Psi_{-1}=v^{1/2} y(v)$, and $\rm z=\frac{4}{m^2} v^{3/2}$, 
turns for $\rm p=1$ into a Bessel equation with the general solution

\begin{equation}
\rm \Psi_{-1}=\left( m^2 A^2 - \kappa  \right)^{\frac{1}{2}}\left[a_0 
J_{\frac{1}{3}}(z) + b_0 J_{-\frac{1}{3}} (z)  \right] ,\qquad where 
\qquad z=\frac{4}{m^2}\left[m^2 A^2 - \kappa\right]^{3/2}
\end{equation}
where $\rm a_0$ and $\rm b_0$ are superposition constants.

\item{} $\rm m^2 < 0$, 

The differential equation for this subcase is

\begin{equation}
\rm -A \Psi^{\prime \prime}_{-1} +p \Psi^{\prime}_{-1} + 144\kappa A^3
\left( |m^2| A^2 + \kappa  \right)  \Psi_{-1}=0 .
\end{equation}
For $\rm p=1$ we have a modified Bessel equation with the general solution

\begin{equation}
\rm \Psi_{-1}=\left( |m^2| A^2 + \kappa  \right)^{\frac{1}{2}}
\left[a_1 I_{\frac{1}{3}}(z)
+ b_1 K_{\frac{1}{3}} (z)  \right] ,\qquad where \qquad 
z=\frac{4}{|m^2|}\left[|m^2| A^2 + \kappa\right]^{3/2},
\end{equation}
where $\rm a_1$ and $\rm b_1$ are superposition constants.

\item{} For $\rm m^2=0$, the differential equation for this situation is

\begin{equation}
\rm -A \Psi^{\prime \prime}_{-1} +p \Psi^{\prime}_{-1} 
+ 144\kappa A^3  \Psi_{-1}=0
\end{equation}
which for $\kappa=1$, has as solution the modified Bessel functions
of order $\rm \nu=\frac{1+p}{4}$
\begin{equation}
\rm \Psi_{-1}=A^{2\nu}\left[A_0 I_\nu(6A^2)
+ B_0 K_\nu (6A^2)  \right], 
\end{equation}
where $\rm A_0$ and $\rm B_0$ are superposition constants, while for
$\rm \kappa=-1$, the solution become to be the ordinary Bessel
functions

\begin{equation}
\rm \Psi_{-1}=A^{2\nu}\left[A_1 J_\nu(6A^2)
+ B_1 Y_\nu (6A^2)  \right], 
\end{equation}
where $\rm A_1$ and $\rm B_1$ are superposition constants.
\end{enumerate} 

\subsection{ Inflationary like scenario,  $\gamma= -\frac{1}{3}$}

In this particular case, the WDW equation becomes
\begin{equation}
\rm A \frac{d^2 \Psi}{dA^2} - p \frac{d\Psi}{dA} - 48 A^3 \left(\Lambda A^2
- g \right)\Psi = 0, \qquad with \qquad g=8\pi G M_{-1/3} - 3\kappa 
\end{equation}
which, after the substitutions $\rm v=\Lambda A^2 - g$, 
$\rm \Psi=v^{1/2} y(v)$, and $\rm z=\frac{4\sqrt{3}}{3\Lambda} v^{3/2}$, 
yields for ${\rm p=1}$ a Bessel equation with the general solution

\begin{equation}
\rm \Psi(A) = \left[\Lambda A^2 - g \right]^{1/2}\, \left\{
C_0 I_{\frac{1}{3}}\left(\frac{4\sqrt{3}}{3\Lambda} \left[\Lambda A^2 - 
g \right]^{3/2}  \right) + 
C_1 K_{\frac{1}{3}}\left(\frac{4\sqrt{3}}{3\Lambda} \left[\Lambda A^2 - 
g \right]^{3/2} \right) \right\}\, .
\end{equation}
where $\rm C_0$ and $\rm C_1$ are superposition constants.

\subsection { Dust case  ($\gamma =0$, $\rm \kappa=0$)}
For this case the WDW equation is
\begin{equation}
\rm \rm A \frac{d^2 \Psi}{dA^2} - p \frac{d\Psi}{dA} + 48 A^2 
\left(-\Lambda A^3 + 8\pi G M_0 \right)\Psi = 0,  
\end{equation}
Making the transformations 
$\rm z=\frac{8}{3}\sqrt{3\Lambda} A^3$ and
$\rm \Psi=e^{-\frac{1}{2}z} u(z)$, one gets for $\rm u(z)$ the confluent 
hypergeometric equation
\begin{equation}
\rm z \frac{d^2 \,u}{dz^2} + \left(\gamma - z \right) \frac{d u}{dz} 
- m \, u =0
\end{equation}
where $\rm m=\frac{2-p}{6}- \frac{16 \pi G M_0}{\sqrt{3\Lambda}}$ and 
$\rm \gamma=\frac{2-Q}{3}$. The linear independent solutions are 
(Gradshteyn and Ryzhik, 1980)
\begin{eqnarray}
{\rm u_1(z)}&=&{\rm  _1F_1\left(m,\gamma;z\right)} \label{u1} \\
{\rm u_2(z)}&=&{\rm z^{1-\gamma} \, _1F_1\left(m-\gamma+1,2-\gamma;z\right)}
\label{u2} 
\end{eqnarray}
where $\rm _1F_1$ is the degenerate hypergeometric function.

Thus, the exact solution for $\Psi$ becomes
\begin{equation}
{\rm \Psi(A)}={\rm    e^{-\frac{1}{2}z} \, \left[ A_0\, u_1(z)
+ B_0  \, u_2(z) \right]} ,\\
\end{equation}
where $\rm A_0$ and $\rm B_0$ are superposition constants, $\rm u_1$ and $\rm u_2$ are the functions given in (\ref{u1})  and
(\ref{u2}), respectively. 
\subsection{ Stiff fluid case  ($\gamma =1$, $\kappa=0$)}
The WDW equation 
\begin{equation}
\rm \rm A \frac{d^2 \Psi}{dA^2} - p \frac{d\Psi}{dA} + 48 \left(-\Lambda A^5
+ 8\pi G M_1 A^{-1} \right)\Psi = 0,  
\end{equation}
has the following exact solution
\begin{equation}
\rm \Psi(A)=A^{\frac{1+p}{2}} \, 
Z_\nu \left(\frac{4\sqrt{-3\Lambda}}{3}\, A^3 \right), \qquad with \qquad 
\nu=\frac{1}{3}\sqrt{\left(\frac{1+p}{2}\right)^2 -384 \pi\,G\, M_1} \, .
\end{equation}
The exact expression  of the wave function depends on the sign of the
cosmological constant and the order $\nu$: 
\begin{itemize}
\item{} $\Lambda >0$ and $\nu$ real, the functions $\rm Z_\nu$ become the 
modified Bessel functions, either $\rm I_\nu$ or $\rm K_\nu$, depending on 
the boundary conditions.
\item{} $\Lambda <0$ and $\nu$ real, the functions $\rm Z_\nu$ turn into the 
ordinary Bessel functions, either $\rm J_\nu$ or $\rm Y_\nu$, depending on 
the boundary conditions.
\item{} $\Lambda >0$ and $\nu$ pure imaginary, the functions $\rm Z_\nu$ 
become the modified Bessel functions  of pure imaginary order
(Dunster, 1990), 
either $\rm I_\nu$ or $\rm K_\nu$, depending on 
the boundary conditions.
\item{} $\Lambda <0$ and $\nu$ pure imaginary, the functions $\rm Z_\nu$ 
turn into the ordinary Bessel functions of pure imaginary order 
(Dunster, 1990),
either $\rm J_\nu$ or $\rm Y_\nu$, depending on 
the boundary conditions.
\end{itemize}

\section{The classical behaviour from WKB regime}
Interesting results can be obtained at the level of 
WKB method if one   
performs the transformation $\rm \Pi_A\rightarrow \frac{d \Phi}{d A}$. Then, 
(\ref{hamiltoniana}) becomes 
the Einstein-Hamilton-Jacobi
equation, where $\Phi$ is the superpotential function that is related 
to the physical potential under consideration.  

Introducing this ansatz in (\ref{hamiltoniana}) we get 

\begin{equation}
\rm H = \frac{1}{24A}\left[\left(\frac{d\Phi}{dA}\right)^2 +144\kappa A^2 
+ 48 \Lambda A^4 -384\pi G  M_\gamma A^{-3\gamma +1}\right]= 0,
\end{equation}
thus, we obtain
\begin{equation}
\rm \frac{d\Phi}{dA} =\sqrt{-48 \Lambda A^4 + 384 \pi G  M_\gamma 
A^{-3\gamma+1} -144\kappa A^2 }
\label{potentia}
\end{equation}
the superpotential $\Phi$ has the following form
\begin{equation}
\rm \Phi = \pm\int \sqrt{-48 \Lambda A^4 + 384 \pi G  M_\gamma A^{-3\gamma+1}
-144\kappa A^2 }\,dA,
\label{phi}
\end{equation}
This integral can be solved for particular cases of the $\gamma$ parameter. 

Now, employing the equations (\ref{wkb},\ref{momento}) and (\ref{potentia})
we will obtain the evolution for the scale factor.
The classical equation of motion is
\begin{equation}
\frac{12 A}{N} \frac{dA}{dt} = \Pi_A= 12 A \sqrt{-\frac{1}{3} \Lambda A^2 + 
\frac{8}{3} \pi G  M_\gamma A^{-3\gamma-1} -\kappa },
\label{classica}
\end{equation}
in term of the ``cosmic time'' $\tau$ defined by $\rm d \tau = N(t) dt$, the
equation (\ref{classica}) is read how (when we choose the gauge N(t)=1, this
cosmic time is the physical time t).
\begin{equation}
\rm d \tau = \frac{dA}{\sqrt{-\frac{1}{3} \Lambda A^2 + 
\frac{8}{3} \pi G  M_\gamma A^{-3\gamma-1} -\kappa }}.
\label{maestra}
\end{equation}

\subsection{Inflationary scenario, $\gamma=-1$}
The equation (\ref{maestra}) for the inflation regime is written as
\begin{equation}
\rm  \tau - \tau_0= \int_0^A \frac{dx}{\sqrt{(-\frac{1}{3} \Lambda  + 
\frac{8}{3} \pi G  M_\gamma) x^2 -\kappa }} = \frac{1}{\sqrt{a_1}} 
\ln\left[ A + \sqrt{A^2 - a_2^2} \right] 
\label{inflation}
\end{equation}
where $\rm a_1=-\frac{1}{3} \Lambda  + \frac{8}{3} \pi G  M_\gamma$
and $a_2^2=\frac{\kappa}{a_1}$.

The inverse of the equation (\ref{inflation}) us give the following
structure for the scale factor $\rm A(\tau)$
\begin{equation}
\rm A(\tau) = \frac{1}{2} \left[ e^{\sqrt{a_1} (\tau -\tau_0)} +
a_2^2 \, e^{-\sqrt{a_1} (\tau -\tau_0)} \right].
\end{equation}
The scale factor $\rm A(\tau)$ will have a inflationary behaviour only if the
parameter $\rm a_1\gg 1$, then the cosmological constant have the restriction
value $\rm \Lambda < 3\left( \frac{8}{3}\pi G M_1 - 1   \right)$. If the
parameter $\rm 0<a_1 <1$, the scale factor vanishes very fast.

For other scenarios and particular values in $\kappa$ and $\Lambda =0$, we
have the following:

\begin{enumerate}
\item{} For the dust case  and $\kappa=0$,  we obtain the 
well-known solution $\rm A \propto t^{2/3}$ for N=1.
 \item{} For $\gamma=1$ and $\kappa=0$, the scale factor have the traditional
solution $\rm A \propto t^{1/3}$ for N=1.
\item{}For radiation case and $\kappa=-1$, the behaviour is
\begin{equation}
\rm A(\tau)= \sqrt{4\left( \tau- \tau_0 \right)^2 - \frac{8}{3}\pi G 
M_{\frac{1}{3}} }.
\end{equation}
For the case of $\kappa=0$, $\rm A \propto t^{1/2}$ for N=1.
\end{enumerate}

On the other hand, the master equation (\ref{maestra}) can be solved for 
any $\gamma$ if 
the cosmological constant vanish, defining $\rm d\tau = A^{3\gamma +2} dT$, 
having the following general solution in the time T
\begin{equation}
\rm A(T) = \left(\frac{3}{8\pi G M_\gamma} \right)^{\frac{1}{3\gamma+1}}
\left[ \left\{\frac{4}{3}\pi G M_\gamma (3\gamma+1) \left(T - T_0\right)
 \right\}^2 + \kappa \right]^{\frac{1}{3\gamma + 1}}
\end{equation} 

The classical solutions obtained by this method perhaps coincide with these
obtained solving the Einstein field equation, but is not possible to
know if the quantum universe remains quantum forever for some case, due to
the fact that the WDW equation is not time dependent.

\section{Conclusions}
We discussed from  the point of view of simple models of 
minisuperspace quantum cosmology.  
We found that the wave functions of the WDW equations in question
are mostly Bessel functions. For particular values of
the parameter p that measures the ambiguity of the factor ordering problem
we obtained analytical results for a some scenarios of 
the universe possessing a cosmological constant. Only in the case of stiff 
fluid, the cosmological constant can be positive or negative. In the
other cases, it can be only positive for physical reasons related to the 
scale factor or the potential under consideration.
It is well known that the Wheeler-DeWitt cosmological equation is not an 
evolution equation and therefore the associated
quantum states do not evolve in time. A possible way out of this difficulty 
could be to connect some parameters of the `quantum' WDW 
solutions with classical Einstein ones by phenomenological restrictions 
imposed on the superpotential function as we did in this work. Using this
method, we find the classical behaviour for the scale factor, that are
analogous to those found solving the Einstein equations.

\bigskip
\noindent {\bf Acknowledgments}\\
We  thank H.C. Rosu  for critical reading of the manuscript.

\bigskip\bigskip
\noindent References

\noindent Bojowald, M. (2002). {\it Physical Review Letters} {\bf 89} 
261301,	gr-qc/0206054.

\noindent Christodoulakis, T. and Zanelli, J. (1984), 
{\it Physics Letters A} {\bf 102}, 227; (1984), {\it Physical Review D} 
{\bf 29}, 2738. 

\noindent Dunster, T. M. Siam Journal Mathematical. Anal. {\bf 21(4)}, 995 (1990). 

\noindent Gibbons, G. W. and  Grishchuk, L. P. (1989). 
{\it Nucluar Physics B} {\bf 313}, 736. 

\noindent Gradshteyn, I. S. and  Ryzhik, I. M. (1980), 
{\it Table of Integrals, Series, and Products}, (Academic Press), page 1059

\noindent Halliwel, J. A. (1991). in S. Coleman, J.B. Coleman, 
T. Piran, S. Weinberg (Eds), {\it Quantum cosmology and baby universes}, (World
Scientific, Singapore).

\noindent Hartle, J. and  Hawking, S. W. (1983). {\it Physical  Review D}
 {\bf 28}, 2960.

\noindent Isham, C. J. (1992). gr-qc/9210011

\noindent Lidsey, J. E. and  Moniz, P. V. (2000) {\it Classical
and  Quantum Gravity}{\bf 17}, 4823.

\noindent Pazos, E. (2000) {\it Aplicaci\'on del 
formalismo lagrangiano ADM a un modelo cosmol\'ogico}, Univ. de San Carlos
de Guatemala, BS thesis in Spanish. 

\noindent  Ryan Jr., M. (1972). {\it Hamiltonian Cosmology}, 
(Springer Verlag).

\noindent Schutz, B. F. (1970). {\it Physical Review D} {\bf 2}, 2762;
(1971). {\it Physical  Review D} {\bf 4}, 3559.

\noindent  Tipler, F. J. (1986). {\it Physics Report} {\bf 137}, 231. 

\end{document}